\documentclass[aps,prl,tightenlines,onecolumn,groupedaddress,superscriptaddress,showpacs]{revtex4}
\usepackage{amssymb}
\usepackage{color,graphicx}
\usepackage{lmodern}
\usepackage{amsmath}
\usepackage{amsbsy}
\usepackage{amsthm}
\usepackage{float}
\usepackage{bbm}
\usepackage{bm}
\usepackage{graphicx}

\begin{document}
\title{Position-Momentum Uncertainty Relation for an Open Macroscopic Quantum System}

\author{Hamid Reza Naeij}
\email[]{naeij@ch.sharif.edu}
\affiliation{Research Group on Foundations of Quantum Theory and Information,
Department of Chemistry, Sharif University of Technology
P.O.Box 11365-9516, Tehran, Iran}
\author{Afshin Shafiee}
\email[Corresponding Author:~]{shafiee@sharif.edu}
\affiliation{Research Group on Foundations of Quantum Theory and Information,
Department of Chemistry, Sharif University of Technology
P.O.Box 11365-9516, Tehran, Iran}
\affiliation{School of Physics, Institute for Research in Fundamental Sciences (IPM), P.O.Box 19395-5531, Tehran, Iran}

\begin{abstract}
 The macroscopic quantum systems are considered as a bridge between quantum and classical systems. In this study, we explore the validity of the original Heisenberg position-momentum uncertainty relation for a macroscopic harmonic oscillator interacting with environmental micro- particles. Our results show that, in the quasi-classical situation, the original uncertainty relation does not hold, when the number of particles in the environment is small. Nonetheless, increasing the environmental degrees of freedom removes the violation bounds in the regions of our investigation.
\end{abstract}
\pacs{03.65.-w, 03.65.Ta, 03.65.Yz}
\maketitle
\textbf{keywords}: Open macroscopic quantum system, Harmonic environment, Heisenberg uncertainty relation.

\section{Introduction}
Macroscopic quantum systems (MQSs) has been broadly studied in literature since 1980 (see, e.g., \cite{leggett0, Garg,leggett2, Taka, Caldeira3}). It is commonly believed that $\it{macroscopic}$ refers to those situations where a large number of particles are involved, or to be more precise, it holds for a system when the number of dynamical degrees of freedom is large \cite{Taka}. The assumption of macro-realism indicates that a MQS with two or more macroscopically distinct states availabel to it, is at all times in one or the other of these states, independent of observation. This assumption is called macroscopic definiteness \cite{Garg}. This may affect some fundamental features of a MQS, specially when the system is in interaction with the environmnet, because the physical parameters of the latter could relatively control the state definitness of the former by affecting the properties of the system in an integrated context.

Moreover, the study of the open quantum systems is an important subject in physics community due to its various applications e.g., in quantum optics, quantum information, solid-state physics, $\it{etc}$. In a general approach, the study of such systems is based on their dissipative and dephasing behaviors \cite{Taka}. This is conventionally pursued by three approaches: a) introducing  an effective Hamiltonain, b) solving a quantum master equation, and c) using a quantum Langevin equation. The last two approaches are both based on modeling the system interacting with a thermal bath, while the first one is described by a time-dependent or non-Hermitian Hamiltonian, which would lead to a dissipative dynamic equation \cite{Yang1, Schlos1, joos}.

Quantum mechanical interference effects may be washed out by the influence of the environment. This can happen regardless of the system in question being macroscopic or not. The effect of the environment on some fundamental behaviors of the system such as the validity of Heisenberg Uncertainty Relation (HUR) for a given pair of incompatible observables is an important problem, when other parameters rather than system's own variables are involved. A supposed HUR shows the inherent difference between the classical and the quantum description of states and sets a limit on the precision of incompatible quantum measurements. As is well-known, the original mathematical form of HUR for the canonical pair of operators including the position $\hat{x}$ and the momentum $\hat{p_x}$ with $[\hat{x} , \hat{p_x}]=i\hbar$, is $\Delta{x^2} \Delta{p_x^2}\geq\hbar^2/4$.

Many works have been done to explain the domain of validity of HURs. Anastopoulos and Halliwell studied the validity of HUR in a class of quantum Brownian motion models consisting of a particle moving in a potential $V (x)$, coupled to an environment of harmonic oscillators in a thermal state \cite{Anastopoulos}. Ozawa  investigated the universally valid reformulation of the HUR as noise and disturbance in measurement. He showed that the HUR lower bound for the noise-disturbance product is violated even by a nearly nondisturbing, precise position measuring instrument \cite{ozawa}. The violation of HUR for two coupled oscillators is also suggested by McDermott and Redmount  \cite{McDermott}. Moreover, there have been proposed other works on HUR and its violation in theoretical and experimental contexts \cite{ozawa1, ozawa2, werner, Busch, Rozema, Sulyok,Tirandaz}.

In our study, instead of using a master equation approach, applied to describe the dynamics of the system under decoherence, we extend a previously known method \cite{McDermott}, in which without any approximation, the stationary states of a macroscopic quantum harmonic oscillator linearly coupled to an environment containing micro-oscillating  degrees of freedom are obtained analytically. Supposing the conditions under which the proposed open system could be treated quasi-classically, we calculate the limits of the original uncertainty relation and the vioilation range of it.

The paper is organized as follows. In section 2, the linearly coupled harmonic-environment model for the system is introduced and analyzed. Then, the stationary states are exactly evaluated for various numbers of the particles of the environment. In section 3, the position and the momentum expectation values of the system are calculated for the corresponding model of the environment, including different number of interacting particles. The violation ranges of the position-momentum HUR are illustrated and discussed, afterwards. Finally, the results are discussed in the conclusion part.


\section{A typical form of the open quantum system: Harmonic model of the environment}

Here, we assume that the central system is a quantum-type harmonic oscillator. The entire system is composed of the system and the environment, in which the latter is supposed to be a collection of micro-oscillators.

The total Hamiltonian could be rewritten as the sum of the Hamiltonian of the system $\hat{H}_s$, the environment $\hat{H}_e$ and the interaction $\hat{H}_{se}$:
\begin{equation}
\label{eq1}
\hat{H}=\hat{H}_s+\hat{H}_e+\hat{H}_{se}
\end{equation}
The total Hamiltonian can be defined as \cite{Taka}:
\begin{equation}
\label{eq2}
\hat{H}={\hat{H}_S}+{\sum_\alpha}\Big[ \frac{\hat{p}^2_\alpha}{2m_\alpha}+\frac{{m_\alpha}{\omega}^2_\alpha}{2}\big(\hat{{x}_\alpha}-f_\alpha(\hat{R})\big)^2\Big]-\frac{1}{2}\sum_\alpha\hbar\omega_\alpha
\end{equation}

\noindent where ${\alpha}$ varies from 1 to $N$, which $N$ is the total number of the environmental oscillators. In relation (2), $ m_\alpha $, $ x_\alpha $, $ p_\alpha $ and $ \omega_\alpha $ denote the mass, the position, the momentum and the frequency of the  environmental particles, respectively. We have also assumed that the system is $\it{linearly}$ coupled to the environment, as is clear from the term $\big(\hat{{x}_\alpha}-f_\alpha(\hat{R})\big)^2$ where $f_\alpha(\hat{R})$ is an arbitrary spatial function of $\hat{R}$ for the system. We have also included the term $-\frac{1}{2}\sum_\alpha\hbar\omega_\alpha$ for latter convenience.

Now, let us begin with the way one can introduce dimensionless parameters for an arbitrary quantum sysytem \cite{Taka}. To do this, we define the characteristic parameters for length $R_0$ and energy $U_0$ as constant units of length and energy, respectively. Subsequently, for a particle of mass $M$, one can define the characteristic time as $\tau_0=R_0 / (U_0/ M)^\frac{1}{2}$. Since, $U_0$ acts like the kinetic energy of a given system, the unit of momentum could be defined as  $P_0=(U_0 M)^\frac{1}{2}$.  Correspondingly, the conjugate variables $q$ and $p$ are defined as $q=R/R_0$ and  $p=P/P_0$. 

In the dimensionless form, one can use the following classes of Hamiltonian for the system, the environment and the system-environment interaction, respectively:
\begin{equation}
\label{eq3}
{\hat{H}_s}=\frac{\hat{p}^2}{2}+V(\hat{q})
\end{equation}
\begin{equation}
\label{eq4}
\hat{H}_e={\sum_\alpha}\Big[ \frac{\hat{p}^2_\alpha}{2}+\frac{\omega^2_\alpha}{2}{\hat x^2_\alpha-}\frac{\bar h\omega_\alpha}{2}\Big]
\end{equation}
\begin{equation}
\label{eq5}
\hat H_{se}=-\sum_\alpha{\omega}^2_\alpha f_\alpha(\hat q)\hat x_\alpha+\frac{1}{2}\sum_\alpha \omega^2_\alpha \big( f_\alpha (\hat q)\big) ^2
\end{equation}

\noindent  The canonical relations are $[\hat{q},\hat{p}]=i\bar h$ and $[\hat{x}_\alpha,\hat{p}_\beta]={i\bar h}\delta_{\alpha\beta}$, where
 \begin{equation}
 \label{eq6}
 \bar h=\frac{\hbar}{{P_0}{R_0}}
 \end{equation}

 As one can see in (4) and (6), instead of Planck constant $\hbar$, a new dimensionless parameter $\bar h$ appears which rates quantitavely the quantum nature of the system. Strictly speaking, the situation in which one gets $\bar h\ll1$, the system behaves quasi-classically. The values of $\bar h$ between 0.01 to 0.1 are fair enough to show the macroscopic disposition of the proposed system \cite{Taka}. In many applications, $R_0$ is defined as characteristic length of resonance between left and right counterparts of a double-well potential. So, $\bar h$ in (6) could be also written as:
\begin{equation}
\label{eq7}
\bar h=\lambdabar_0/R_0
\end{equation}
\noindent where $\lambdabar_0=\lambda_0 / 2\pi$. Here, $\lambda_0$ denotes the characteristic wavelength of the central system. For a MQS, $\lambda_0$ is too small compared to $R_0$ which is nearly a fixed value for known models of potential. Thus, regarding the quasi-classical systems, the condition $\bar h<0.1$ seems suitable for our future purposes. Notice that the macroscopically feature of the system independent of whether the system is open or closed.

Hereafter, for simplicity, we assume that $ f_\alpha(q)={\gamma_\alpha}q$ where $ 0<\gamma_\alpha<1$ denotes the strength of the coupling between the system and the environment. Using the above assumption, one reaches that $\hat H_{se}$ is linear both to $ \hat x_\alpha$ and $\hat q$. So, the model is called $\it{bilinear}$. The bilinear assumption is most fitted for the case of limited numbers of the surrounding particles. As $N$ increases, interactions between the particles of the environment grow and the bilinear assumption could not be estabilished. For diminished degrees of freedom, however, the approach in this section is proper enough to show the suppression of the quantum behavior of the system due to the cogency of the model.

Regarding the bilinear condition, the total form of the Hamiltonian can be provided as:
\begin{align}
\label{eq8}
H=\frac{p^2}{2}+V_1(q)+\sum_\alpha\Big[\frac{p^2_\alpha}{2}+\frac{\omega^2_\alpha}{2}x^2_\alpha(1-\gamma_\alpha)-\frac{\bar h\omega_\alpha}{2}\Big]+\frac{1}{2}\sum_\alpha \gamma_\alpha\omega^2_\alpha(x_\alpha-q)^2
\end{align}
where $V_1(q)=V(q)-\frac{1}{2}\big( \sum_\alpha{\gamma_\alpha}(1-\gamma_\alpha)\omega^2_\alpha\big) q^2$.

\noindent Here, $V(q)=\frac{1}{2}\omega_0 ^2 q^2$ is the potential and $ \omega_0 $ is the vibrational frequency of the central system, respectively. This shows that the particle feels the efficient potential $V_1(q)$. Moreover, the main coupling is between each environmental oscillator $\alpha$ with spring constant $(1-\gamma_\alpha)\omega^2_\alpha$ and the central particle having the spring constant $\gamma_\alpha\omega^2_\alpha$.
In the next part, we obtain the stationary eigenfunctions of the total Hamiltonian for the ground state of the entire system.

\subsection{Analytical solution of Schrodinger equation}
Considering the bilinear assumption, the total Hamiltonian can be rewritten as:
\begin{equation}
\label{eq9}
H=\frac{p^2}{2}+\frac{1}{2}\omega^2 q^2+\sum_\alpha\Big[ \frac{p^2_\alpha}{2}+\frac{\omega^2_\alpha}{2}x^2_\alpha-\omega^2_\alpha\gamma_\alpha q x_\alpha\Big]
\end{equation}
where $\omega=[\omega_0 ^2+\sum_\alpha \omega^2_\alpha \gamma^2_\alpha]^\frac{1}{2}$.

The total Hamiltonian in (9) reads as the sum of the Hamiltonian indexing $\alpha$, which can be defined as

\begin{equation}
\label{eq10}
H_\alpha=\frac{p'^2}{2}+\frac{1}{2}\omega^2 q'^2+ \frac{p^2_\alpha}{2}+\frac{1}{2} \omega^2_\alpha x
^2_\alpha-\omega^2_\alpha \gamma_\alpha q' x_\alpha
\end{equation}
where $p'=p/N , q'=q/N$ and the last term indicates interaction between the environmental particle and the central system.

Now, we decouple the Hamiltonian by using the rotation of the position coordinates $(q',x_\alpha)$ and the momentums $(p',p_\alpha)$ to define new position and momentum coordinates, $(x_+,x_-)$ and $(p_+, p_-)$, respectively:
\begin{equation}
\label{eq11}
\begin{pmatrix}

x_+  \\
x_-
\end{pmatrix}
\begin{matrix}\\\mbox{}\end{matrix}
=\begin{pmatrix} cos\theta & sin\theta \\ -sin\theta & cos\theta \end{pmatrix}
\begin{pmatrix} q' \\
x_\alpha
\end{pmatrix}
\end{equation}
\noindent and
\begin{equation}
\label{eq12}
\begin{pmatrix}

p_+  \\
p_-
\end{pmatrix}
\begin{matrix}\\\mbox{}\end{matrix}
=\begin{pmatrix} cos\theta & sin\theta \\ -sin\theta & cos\theta \end{pmatrix}
\begin{pmatrix} {p'} \\
{p}_\alpha
\end{pmatrix}
\end{equation}

\noindent Under the rotation, the kinetics energy part of the Hamiltonian in (10) remaines invariant. Thus, decoupling of the Hamiltonian is achieved by diagonalizing the potential energy. In the following condition,

\begin{equation}
\label{eq13}
\theta=\frac{1}{2}arctan\Big[ \frac{\omega'^2_\alpha}{\omega^2 -\omega^2_\alpha}\Big]
\end{equation}
the rotations transform the Hamiltonian to
\begin{equation}
\label{eq14}
H_\alpha=\frac{p_{+\alpha}^2}{2}+\frac{1}{2} \omega_{+\alpha}^2 x_{+\alpha}^2+\frac{p_{-\alpha}^2}{2}+\frac{1}{2} \omega^2_{-\alpha} x_{-\alpha}^2
\end{equation}
where $H_\alpha =H_{+\alpha}+H_{-\alpha}$.

\noindent Here, we define:
\begin{equation}
\label{eq15}
\omega_{+\alpha}=\big[ \omega^2 \cos^2\theta+\omega^2_\alpha \sin^2\theta+\omega'^2_\alpha \sin\theta\cos\theta\big]^\frac{1}{2}
\end{equation}
\begin{equation}
\label{eq16}
\omega_{-\alpha}=\big[\omega^2 \sin^2\theta+\omega^2_\alpha \cos^2\theta-\omega'^2_\alpha \sin\theta\cos\theta\big]^\frac{1}{2}
\end{equation}
\noindent where $\omega'_\alpha=i ( 2{\omega^2_\alpha}{\gamma_\alpha})^\frac{1}{2}$. Now, we assume that:

{$\mathbb{A}$) For all particles of the environment, $\omega_\alpha$ and $\gamma_\alpha$ are nearly the same.}

Consequently, we define $tan2\theta=c$ in (13). If $c>0$, then we should have $\omega^2<\omega^2_\alpha$. \noindent This means that
\begin{equation}
\label{eq17}
\omega^2_0<\omega^2_\alpha(1-\gamma^2_\alpha N)
\end{equation}
\noindent where $\omega^2_0=\omega^2-\omega^2_\alpha\gamma^2_\alpha N$. Yet, in (17), it is necessary that $(1-\gamma^2_\alpha N)>0$, or $N\gamma^2_\alpha <1$. So, the number of particles in the environment should restrict the strength of interaction $\gamma_\alpha$, which is not a legitimate condition. If we take $c<0$, it will result that
\begin{equation}
\label{eq18}
\lambda^2_0(1-\gamma^2_\alpha N)<\lambda^2_\alpha
\end{equation}
\noindent where $\lambda_\alpha$ is the wavelength of the environmental particles. For both small values of $N$ and $\gamma_\alpha$ (so that $1-\gamma^2_\alpha N\approx1$), one concludes from (18) that
\begin{equation}
\label{eq19}
\bar h=\lambdabar_0/R_0<\lambdabar_\alpha/R_0
\end{equation}
\noindent which guarantees the quasi-classical behavior of the central system. This is because for a MQS, the characteristic wavelength of the macro-system $\lambda_0$ is much smaller than the corresponding wavelength $\lambda_\alpha$ of the micro-particles of the environment. We choose $0.01<\bar h=\lambdabar_0/R_0<0.1$ to reach the definite bound of $\bar h$ for a MQS, as mentioned before. Also, in (18), one can notice that $N\gamma^2_\alpha >(1-\lambda^2_\alpha / \lambda^2_0)$. If one assumes that $\lambda_\alpha / \lambda_0<1$ \big(contrary to (19)\big), the values of $N$ and $\gamma_\alpha$ will be again restricted to a positive constant value which is not reasonable, since they are independent parameters. So, for any value of $N$ and $0<\gamma_\alpha<1$, $\lambda_\alpha / \lambda_0>1$, the condition of (19) is compelling. This shows that under the assumption $\mathbb{A}$, no microscopic quantum system could be a good candidate for our model.The key point is that the emergence of classicality, here, is due to the conditions that the system interacts with the environment.

From the Hamiltonian (14), wave function in position space for the ground state of the total system is obtained as:
\begin{align}
\label{eq20}
\psi_0(x_{+\alpha} , x_{-\alpha})=\prod_{\alpha=1}^{N}\Big(\frac{{\omega_{+\alpha}} {\omega_{-\alpha}}}{{\pi^2\bar h^2}}\Big)^\frac{1}{4}\exp\Big(\frac{-{\omega_{+\alpha}}{x_{+\alpha}^2}}{2\bar h}\Big)\exp\Big(\frac{-{\omega_{-\alpha}}{x_{-\alpha}^2}}{2\bar h}\Big)
\end{align}

In the next section, we will calculate position and momentum expectation values of the system coupled to $N=2,3$  environmental oscillators. Limited number of the environmental particles are taken here to support the pragmatic facet of assumption $\mathbb{A}$.

\section{Position and momentum expectation values of an open MQS}
\subsection{$\textbf{N=2}$}

Considering the definitions $a=\sin\theta$, $b=\cos\theta$ and $c=\tan2\theta <0$, plus-minus position coordinates of the two particles of the environment can be written as:
\begin{align}
\label{eq21}
x_{+\alpha}&=bq'+a{x}_\alpha \nonumber\\
x_{-\alpha}&=-aq'+b{x}_\alpha
\end{align}
\noindent where $\alpha=1,2$. In this case, the normal ground state wave function is given by:
\begin{align}
\label{eq22}
\psi_0(x_{+1} , x_{-1},x_{+2},x_{-2})= \Big(\frac{{\omega_{+1}} {\omega_{-1}}{\omega_{+2}}{\omega_{-2}}}{{\pi^4\bar h^4}}\Big)^\frac{1}{4}
\times \exp\Big(\frac{-{\omega_{+1}}{x_{+1}^2}}{2\bar h}\Big)\exp\Big(\frac{-{\omega_{-1}}{x_{-1}^2}}{2\bar h}\Big)\exp\Big(\frac{-{\omega_{+2}}{x_{+2}^2}}{2\bar h}\Big)\exp\Big(\frac{-{\omega_{-2}}{x_{-2}^2}}{2\bar h}\Big)
\end{align}

The probability distribution for the position coordinate of the system, $P(q)$, is obtained by integrating  the probability density over the spatial coordinates of the environmental oscillators ($x_{1} , x_{2}$). Regarding the relation (21) and using the wave function (22), one gets:
\small
\begin{align}
\label{eq23}
 &P(q)=\frac{1}{\pi \bar h}\Big[\frac{\omega_{+1} \omega_{-1}\omega_{+2}\omega_{-2}}{(a^2\omega_{+1}+b^2\omega_{-1})(a^2\omega_{+2}+b^2\omega_{-2})}\Big]^\frac{1}{2}\nonumber \\
&\times\exp\Big[{\frac{-q^2 [(a^2\omega_{+1}\omega_{+2})(\omega_{-1}+\!\omega_{-2})+(b^2\omega_{-1}\omega_{-2})(\omega_{+1}+\omega_{+2})]}{\bar h(a^2\omega_{+1}+b^2\omega_{-1})(a^2\omega_{+2}+b^2\omega_{-2})}\Big]}
\end{align}
\normalsize
\noindent So, the position expectation values $\langle q\rangle$ and $\langle q^2\rangle$ are respectively calculated as:
\begin{equation}
\label{eq24}
\langle q\rangle =\int_{-\infty}^{+\infty} P(q) q dq=0
\end{equation}
\noindent and
\begin{align}
\label{eq25}
\langle q^2\rangle &= \int_{-\infty}^{+\infty} P(q) q^2 dq\nonumber\\
 &= \frac{1}{2} \sqrt{\frac{\bar h}{\pi}} ({\omega_{+1}} {\omega_{-1}}{\omega_{+2}}{\omega_{-2}})^\frac{1}{2}[(a^2\omega_{+1}+b^2\omega_{-1})(a^2\omega_{+2}+b^2\omega_{-2})]{[(a^2{\omega_{+1}}{\omega_{+2}})(\omega_{-1}+\omega_{-2})}+{(b^2{\omega_{-1}}{\omega_{-2}})(\omega_{+1}+\omega_{+2})]}^\frac{-3}{2}
\end{align}

For evalulating the momentum expectation values, since the Hamiltonian is decoupled in new coordinates, we can use the Fourier transformation to obtain the wave function in the momentum space. This leads to the normal wave function in the momentum space as:
\begin{align}
\label{eq26}
\psi_0(p_{+1} , p_{-1},p_{+2},p_{-2})=\Big(\frac{1}{\pi^4\bar h^4 \omega_{+1}\omega_{-1}\omega_{+2}\omega_{-2}}\Big)^\frac{1}{4}\exp\Big(\frac{-p_{+1}^2}{2\bar h\omega_{+1}}\Big)\exp\Big(\frac{-p_{-1}^2}{2\bar h\omega_{-1}}\Big) \exp\Big(\frac{-p_{+2}^2}{2\bar h\omega_{+2}}\Big)\exp\Big(\frac{-p_{-2}^2}{2\bar h\omega_{-2}}\Big)
\end{align}

The probability distribution for the momentum of the system, $P(p)$, is obtained by integrating  the probability density over the momentum coordinates of the environmental oscillators ($p_{1} , p_{2}$):
\begin{align}
\label{eq27}
P(p)=\frac{1}{\pi \bar h}\big[(a^2\omega_{-1}+b^2\omega_{+1})(a^2\omega_{-2}+b^2\omega_{+2})\big]^\frac{-1}{2}\exp\Big[\frac{-p^2 [a^2(\omega_{-1}+\omega_{-2})+b^2(\omega_{+1}+\omega_{+2})]}{\bar h(a^2\omega_{-1}+b^2\omega_{+1})(a^2\omega_{-2}+b^2\omega_{+2})}\Big]
\end{align}

Finally, the expectation values $\langle p\rangle$ and $\langle p^2\rangle$ for the system are respectively obtained as:
\begin{equation}
\label{eq28}
\langle p\rangle =\int_ {-\infty}^{+\infty} P(p) p dp=0
\end{equation}
\noindent and
\begin{align}
\label{eq29}
\langle p^2\rangle &=\int_{-\infty}^{+\infty} P(p) p^2 dp\nonumber\\
 &=\frac{1}{2} \sqrt{\frac{\bar h}{\pi}} \big[(a^2\omega_{-1}+b^2\omega_{+1})(a^2\omega_{-2}+b^2\omega_{+2})\big]{\big[a^2(\omega_{-1}+\omega_{-2})+b^2(\omega_{+1}+\omega_{+2})\big]}^\frac{-3}{2}
\end{align}

It is important to note that the system is not in a pure state, here. Since the total system is in a stationary state, however, the time evolution of it is not significant. So, the marginal probability distributions (23) and (27) are not time dependent. This is a main difference between Gaussian free wave packets and such states. For the former case, time evolution is crucial and results in the spatial dispersion of the wave packet at later times. This affects the position and momentum variances, discussed in the next section.

\subsection{$\textbf{N=3}$}
Using the similar method described for $N=2$, one gets the following results for the desired expectation values $ \langle q^2\rangle $ and $\langle p^2\rangle$, where $\langle q\rangle =\langle p\rangle=0$ as before. So,
\begin{align}
\label{eq30} 
\langle q^2\rangle = &\frac{1}{2\pi}(\omega_{+1}\omega_{-1}\omega_{+2}\omega_{-2}\omega_{+3}\omega_{-3})^\frac{1}{2}[(a^2\omega_{+1}+b^2\omega_{-1})
(a^2\omega_{+2}+b^2\omega_{-2})(a^2\omega_{+3}+b^2\omega_{-3})]\nonumber\\
 &\times \lbrace a^4(\omega_{+1}\omega_{+2}\omega_{+3})(\omega_{-1}+\omega_{-2}+\omega_{-3})
 +a^2b^2[(\omega_{-1}\omega_{+2}\omega_{+3})(\omega_{-2}+\omega_{-3})
 +(\omega_{+1}\omega_{-2}\omega_{-3})(\omega_{+2}+\omega_{+3})\nonumber\\
 &+(\omega_{+1}\omega_{-1})(\omega_{+2}\omega_{-3}+\omega_{-2}\omega_{+3})] +b^4(\omega_{-1}\omega_{-2}\omega_{-3})(\omega_{+1}+\omega_{+2}+\omega_{+3})\rbrace^\frac{-3}{2}
\end{align}
\noindent and
\begin{align}
\label{eq31}
\langle p^2\rangle= &\frac{1}{2\pi} [(a^2\omega_{-1}+b^2\omega_{+1})(a^2\omega_{-2}+b^2\omega_{+2})(a^2\omega_{-3}+b^2\omega_{+3})] \lbrace a^4[(\omega_{-2}\omega_{-3})+\omega_{-1}({\omega_{-2}}+{\omega_{-3}})]\nonumber\\
&+a^2b^2[\omega_{+2}\omega_{-3}+\omega_{+1}(\omega_{-2}+ \omega_{-3})
+\omega_{-2}\omega_{+3}+\omega_{-1}(\omega_{+2}+\omega_{+3}) ]
+b^4[\omega_{+2}\omega_{+3}+\omega_{+1}(\omega_{+2}+\omega_{+3})]\rbrace^\frac{-3}{2}
\end{align}

\section{Evaluating the violation range of position-momentum Uncertainty relation}

In the dimensionless regime, HUR for the position and the momentum variances is given by:
\begin{equation}
\label{eq32}
\Delta{q^2} \Delta{p^2}\geq\frac{\bar h^2}{4}
\end{equation}

\noindent where the upper bound of $\bar h$ is given by (19), so that we choose $\bar h< 0.1$. Considering the assumption $\mathbb{A}$, the general form of $\omega_{+\alpha}$ and $\omega_{-\alpha}$ in (15) and (16) can be written as:
\begin{equation}
\label{eq33}
\omega_{+\alpha}=\omega_\alpha \big[1-\gamma_\alpha \frac{b}{a}\big]^\frac{1}{2}
\end{equation}
\noindent and
\begin{equation}
\label{eq34}
\omega_{-\alpha}=\omega_\alpha \big[1+\gamma_\alpha \frac{a}{b}\big]^\frac{1}{2}
\end{equation}
\noindent respectively, where $a^2+b^2=1$. We also define $d=\omega_{+\alpha}/ \omega_{-\alpha}$, where we have supposed that $d$ is the same for any particle of the environment. Then, one can obtain the following ranges of the violation of (32) for different values of $N$ as follows.

\noindent For $N=2$
\begin{equation}
\label{eq35}
0.04<\bar h<0.1
\end{equation}
\noindent For $N=3$
\begin{equation}
\label{eq36}
0.06<\bar h<0.1
\end{equation}
\noindent For $N=4$, the similar calculation gives the folllowing result:
\begin{equation}
\label{eq37}
0.08<\bar h<0.1
\end{equation}

Generally, the lower limit of the violation range is obtained as:
\begin{equation}
\label{eq38}
\frac{1}{{N^3}{\pi^{N-1}}}<\bar h^{N-1}
\end{equation}

\noindent As an instance, for $N=3$, the details of calculations are given in Appendix A.

\begin{figure}[H]
\centering
\includegraphics[scale=0.5]{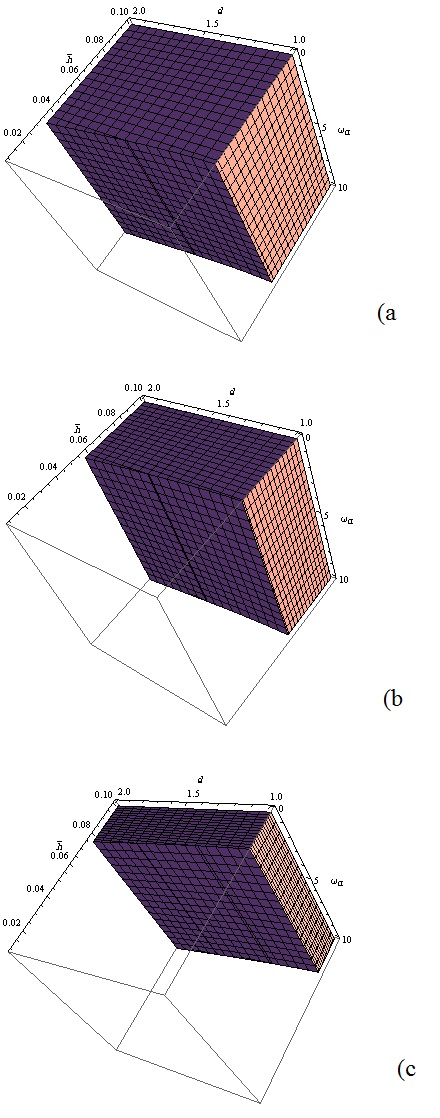}
\caption{The system coupled to a) two, b) three and c) four environmental oscillators. The range of the violation of HUR  (coloned area) is decrased as $N$ increases.}\label{fig1}
\end{figure}

We have drawn  the plots which show the area in which HUR is violated. In these plots, three varying parameters including $\bar h$, the frequency of the environmental particles $\omega_\alpha$ and $d$ are considered.
As is clear in these figures, as the number of the environmental oscillators increases, the area in which the violation of HUR occurs would be smaller (See Fig. \ref{fig1}).

Moreover, it is interesting to see which of the variances has greater effect on the violation of the position-momentum uncertainty relation in our model. For this purpose, we compare the position and the momentum variances of the macroscopic harmonic oscillator which obtained in the previous section, with the closed harmonic oscillator (CHO). 

In the dimensionless regime, the position and the momentum variances of the ground state of a CHO are $\bar h/ 2\omega_0$ and $\bar h \omega_0/ 2$, respectively, where $\omega_0$ is the frequency of the CHO, as defined before. We know that the HUR in (32) is satisfied for CHO.

Let us consider, as an instance, two environmental particles, $N=2$. Using the assumption $\mathbb{A}$ in section 2 and for small values of $\gamma_\alpha$, the position and the momentum variances, (25) and (29) could be simplified to

\begin{equation}
\label{eq39}
\Delta{q^2}=\frac{1}{2^{5/2}}(\frac{\bar h}{\pi \omega_\alpha})^{\frac{1}{2}}
\end{equation}
and
\begin{equation}
\label{eq40}
\Delta{p^2}=\frac{1}{2^{5/2}}(\frac{\bar h \omega_\alpha}{\pi })^{\frac{1}{2}}
\end{equation}

 Thus, one conclude that 
\begin{equation}
\label{eq41}
\frac{\Delta{q^2}}{\bar h/ 2\omega_0}=\frac{1}{2^{3/2}}\frac{\omega_0}{\sqrt{\pi \omega_\alpha \bar h}}>1
\end{equation}
since $\omega_0>\omega^{1/2}_\alpha$. Also, 

\begin{equation}
\label{eq42}
\frac{\Delta{p^2}}{\bar h \omega_0/ 2}=\frac{1}{2^{3/2}}\frac{\sqrt{\omega_\alpha}}{{\sqrt{\pi \bar h}\omega_0}}<1
\end{equation}
for the same reason. So, the uncertainty in position increases for the open system, while its momentum uncertainty decreases. Yet, this result leads to the violation of the HUR in (35). Similar method could be used for $N=3$, with the same result.

Thus, the position variance of the macro-system has more effect on the violation of the position-momentum uncertainty relation. Why does the macro-system show such behavior?

The interaction Hamiltonian between the macro-system and the environmental particles in (5) is position-based \big (see also (9)\big). In other words, the particle $q$ feels the effective potential $V_1(q)$ instead of $V(q)$ and each environmental oscillator $\alpha$ feels the spring constant $(1-\gamma_\alpha )\omega^2_\alpha$, as is obvious in (8). As a result, one cannot interpret $q$ as the center-of-mass position of the central system and $\lbrace x_1, x_2,..., x_N\rbrace$ is indeed the set of degrees of freedom respresenting fluctuations around $q$. This causes the greater values of position variances in (32).

\section{Conclusion}

For a macroscopic quantum system as described by Leggett \cite{leggett0,Garg,leggett2} and others \cite{Taka,Caldeira3}, the assumption of macroscopic definiteness indicates that one can assign distinct states to the macro-system with the property of being in a particular state, even if it is not observed. Yet, it has not been suggested to what extent, if any, this may influence the validity of the HUR, when there is no general form of Heisenberg-type inequalities to include macroscopic systems. Nevertheless, it is a matter of controversy if such inequalities could be valid in transition from quantum to classical situation. For this reason, we guessed that the violation of the position-momentum HUR for an open macroscopic quantum system may occur in certain conditions.

We have instantiated in this paper a macroscopic quantum harmonic oscillator interacting with various numbers of micro-oscillating particles in the environment. Taking into account the value of dimensionless parameter $\bar h$ as an indication of the macroscopic disposition of the system (with an upper bound $\bar h<\lambdabar_\alpha/R_0$), our calculations show that for the quasi-classical systems with $0.04 <\bar h<0.08 $, the macroscopic behavior could be indicated by the violation of position-momentum HUR, when similar limited numbers of the particles of the environment $N$ are in interaction. Such a result could not be predicted  {\it a priori}, specially for small values of $N$. Moreover, our calculations show that the position variances of the macro-system has greater effect than the momentum varinces on the violation of HUR. That is due to the position-position interaction between the macro-system and the environmental particles which affects the real potential that the central system feels.

On the other hand, when $ \bar h>\lambdabar_\alpha/R_0$, and $\lambda_0 >\lambda_\alpha$, we have a micro-system in interaction with the environment. This situation, however, is unfounded under the condition of similarness of the environmental particles. So, the violation of HUR for a microscopic quantum system is beyond our consideration. Also, as the number of the particles of the environment increases, there would appear other factors which could affect the quantum traits of the system. For example, the bilinear assumption does not hold, when many degrees of freedom in the surrounding are involved, or when the internal interactions between the particles of the environment could not be ignored. In such circumstances, the similarity assumption of the environmental particles is denied and  the validity of our result is avoided. For two to four particles, however, the assumptions leading to the violation seem reasonable. So, it is tenable to show such violations with such simple models. This can shed new light on the relationship between the classical trait of a system and its quantum uncertainties illustrated by a proper HUR. Here, we have shown that these two concepts are interrelated.

\appendix*

\section{Appendix A}
\setcounter{equation}{0}
Here, we derive the violation range of uncertainty relation, when the system is coupled to three particles of the environment.
Considering a given violation range as
\begin{equation}
\Delta{q^2} \Delta{p^2}<\frac{\bar h^2}{4}
\end{equation}
One can show that the product of $\langle q^2\rangle\langle p^2\rangle$ in relations (30) and (31) leads to:
\begin{align}
\frac{1}{4\pi^2}\frac{[(a^2b^2+1)d^2+(1-2a^2b^2)d]^3}{[(3a^4d^2+6a^2b^2d+3b^4)(3a^4+6a^2b^2d+3d^2b^4]^\frac{3}{2}}
\end{align}
\noindent where $a^2+b^2=1$ and $d$ is defined. The above relation is simplified to $ 1/4\pi^2(3)^3$, after some mathematical manipulation. This concludes the relation (38) for $N=3$. The same method could be used to prove similar results for other $N$s.

\end{document}